\vspace{30px}
\documentclass{aastex631}
\makeatletter
\let\frontmatter@title@above=\relax
\makeatother
\usepackage{lipsum}% for placeholder text
\usepackage{amsmath} % AMS Math Package
\usepackage{enumitem}
\usepackage{varwidth}
\usepackage{tasks}
\usepackage{amsthm} % Theorem Formatting
\usepackage{amssymb} % Math symbols such as \mathbb
\usepackage{filecontents}
\usepackage{graphicx}
\usepackage{enumitem}
\graphicspath{ {./images/} }

\begin{document}
\vspace*{1cm}
\begin{center}
{\LARGE{$C^{k}$-regular extremal black holes in maximally-symmetric spacetime and the third law of black hole thermodynamics}}
\end{center}

\begin{center}
\Large{Ryan Marin}\\
\bigskip
\small{Primary Advisor: Mihalis Dafermos\\
Secondary Advisor: Frans Pretorius\\
\bigskip
\emph{Department of Physics, Princeton University,\\
Washington Road, Princeton, New Jersey, 08544, USA}}\\
\bigskip
\Large{November 26, 2024}
\end{center}

\begin{abstract}

In this work we extend the proof of Ryan Unger and Christoph Kehle's 2022 work, "Gravitational collapse to extremal black holes and the third law of black hole thermodynamics", to construct examples of black hole formation from regular, one-ended asymptotically flat Cauchy data for the Einstein–Maxwell-charged scalar field system in maximally-symmetric 3+1 dimensional spacetime which are exactly isometric to $dS_{4}/AdS_{4}$ Reissner Nördstrom black holes after a finite advanced time along the event horizon. Furthermore, the apparent horizon coincides with that of a vacuum at finite advanced time. 

This paper exists as an extension of the aforementioned work done by Unger and Kehle which disproves the "third law of black hole thermodynamics" originally posed by Hawking and Bardeen's "The Four Laws of Black Hole Mechanics". We begin with a brief introduction to the history of black hole thermodynamics, tracing the the lineage of the third law of black hole thermodynamics up to Kehle \& Unger's 2022 work; the basepoint for our extension to dS/AdS spacetime. We adapt the machinery from Kehle and Unger to Schwarzschild and Reissner-Nördstrom solutions in dS, AdS space. Then, we study the relevant manifold gluing theory and reprove necessary gluing theorems in $dS_{4}$ \& $AdS_{4}$. Finally, we provide analysis of the third law of black hole thermodynamics in these maximally symmetric spacetimes.
\end{abstract}

\section{\textbf{Introduction}}

The study of black holes as a consequence of \emph{gravitational collapse}, that is a black hole spacetime containing a one-ended Cauchy surface which lies outside of the black hole region, was constructed by Oppenheimer and Snyder [OS39] for the Einstein-massive dust model in spherical symmetry. Naturally, this model was the theoretical framework from which black holes were first studied as thermodynamic objects, where Bardeen, Carter, and Hawking began the analogy of black holes to classical thermodynamics with their proposal of the \emph{four laws of black hole thermodynamics} [BCH73]. In particular, letting the surface gravity $\kappa$ of the black hole take the role of its temperature, they proposed a third law in analogy to “Nernst’s theorem” in classical thermodynamics.
\\

\emph{\textbf{The proposed third law of black hole thermodynamics [BCH73]}: A subextremal black hole cannot become extremal in finite time by any continuous process, no matter how idealized, in which the spacetime and matter fields remain regular and obey the weak energy condition.}
\\
 
This is the origin of the long-standing analogy between black hole thermodynamics and their classical counterparts. It was also served as the groundwork for W. Israel's work on proving the third law of black hole thermodynamics (see: [Isr86; Isr92]), which he, along with others who narrowed the scope of the third law, established as follows:
\\

\emph{\textbf{The third law of black hole thermodynamics [Isr92]} Any continuous process in which the stress-energy tensor of the accretive matter stays bounded satisfies the weak energy condition in a neighborhood of the outer apparent horizon.}
\\

However, Israel's proof had made an incorrect assumption about the nature of apparent horizons. In the context of [Isr92] and [Ung22], the apparent horizon is the set of points in a given spacetime such that all null geodesics converge. In the static case, this corresponds to the definition of an event horizon, however it's dynamical, local description of convergence is more suitable for the dynamical black holes for which the third law applies.

This rigorous definition of the third law of black hole thermodynamics was shown to be incorrect by Kehle and Unger, who utilized manifold gluing techniques to achieve extremality under Israel's definition, achieving $C^{k}$ regular solutions from  regular, one-ended asymptotically flat Cauchy data. The work done by Kehle and Unger is manifestly a disproof of Israel's statement of the third law of black hole thermodynamics embedded in a more general theorem regarding the existence initial regular Cauchy data which evolves into an extremal black hole in finite advanced time. 

Their proof relies on the gluing of two different $C^{k}$ Cauchy data sets along a null cone to glue flat $\mathbb{R}^{3+1}$ Minkowski space to the Schwarzschild solution (Corollary 2.1) and later the an arbitrary Reissner-Nördstrom solution (Corollary 2.2) in finite advanced time. This allows them to product extremal black holes for specific choice of initial and final data sets, but they restrict their work to the case of $\Lambda = 0$\footnote{$\Lambda$ is the cosmological constant, so $\Lambda = 0$ indicates a universe with no intrinsic curvature.}, or a Minkowski background space time. 

In this work we extend much of the theory to de Sitter and Anti-de Sitter spacetime, particularly extending the gluing theorems to both and extremality to AdS space. This work is intended as a note on the work by Kehle and Unger, and as such should be read in conjunction with the original paper. Particularly, this work includes only the modifications required for an analysis in dS/AdS space; much of the derivation/analysis which remains invariant in the case of a nonzero cosmological constant is omitted.
\\

\section{\textbf{Spherically Symmetric EMSCF equations with $\Lambda \neq 0$}}

As in the original work, we will work under the EMSCF equations whose initial conditions will be the Cauchy data:

\[
R_{\mu\nu} - \frac{1}{2}Rg_{\mu\nu} + \Lambda g_{\mu\nu}= 2(T^{EM}_{\mu\nu} + T^{CSF}_{\mu\nu})
\]
\[
\nabla_{\mu}F_{\mu\nu} = 2q\Im(\phi\overline{D_{\nu}\phi})g^{\mu\nu}D_{\mu}D_{\nu}\phi = 0
\]
\[
g^{\mu\nu}D_{\mu}D_{\nu}\phi = 0
\]

Where we define $T^{EM}_{\mu\nu}$ and $T^{CSF}_{\mu\nu}$ in the standard form:

\[
T^{EM}_{\mu\nu} = g^{\alpha\beta}F_{\alpha\nu}F_{\beta\mu} - \frac{1}{4}F^{\alpha\beta}F_{\alpha\beta}g_{\mu\nu}
\]
\[
T^{CSF}_{\mu\nu} = \text{Re}(D_{\mu}\phi\overline{D_{\nu}\phi}) - \frac{1}{2}g_{\mu\nu}g^{\alpha\beta}D_{\alpha}\phi\overline{D_{\beta}\phi}
\]

for a quintuplet ($\mathcal{M}, g, F, A, \phi)$\footnote{The next few sections assume $\Lambda = 0$, as these sections pertain to background leading up to the main theorems of [Ung22].}, where $(\mathcal{M}, g)$ is a $3+1$ dimensional Lorentzian manifold, $\phi$ is a complex scalar field, $A$ is a real one-form, $F = dA$ is a real-valued 2-form, and $D = d + iqA$ is the gauge covariant derivative, $q \in \mathbb{R}/\{0\}$ is a fixed coupling constant for the scalar field $\phi$.

To examine how a non-Minkowski background spacetime will affect the formation of extremal black holes, we must first establish a system of Einstein-Maxwell Scalar field equations in dS/AdS space. Note that we still have the same two parallel transport equations from the condition that 

\[
\nabla^{\mu}F_{\nu\mu} = 0
\]

and thus the Maxwell equations remain unchanged (see [Kom13] for details):

\[
\partial_{\mu}Q = -qr^{2}\Im(\phi\overline{D_{\mu}\phi})
\\
\partial_{\nu}Q = qr^{2}\Im(\phi\overline{D_{\nu}\phi})
\]

Furthermore, we retain the gauge choice $A_{u} = 0$, and thus have

\[
F = \frac{Q\Omega^{2}}{2r^{2}} du \wedge dv
\\
\partial_{v}A_{u} = -\frac{Q\Omega^{2}}{2r^{2}}
\]

For a derivation of equations analogous to those in [Ung22]\footnote{Equations (2.4)-(2.11)}, we reference [Ros23] appendix A. Note the following form of the wave equations (where we set $m=0$ from [Ros23], which will be referenced in further chapters.

\[
\partial_{u}\partial_{v}\phi = -\frac{\partial_{u}\phi\partial_{v}r}{r}-\frac{\partial_{u}r\partial_{v}\phi}{r}+\frac{iq\Omega^{2}Q}{4r^{2}}\phi-iqA_{u}\frac{\partial_{v}r}{r}\phi-iqA_{u}\partial_{v}\phi
\]
\[
\partial_{u}\partial_{v}r = \frac{\Omega^{2}}{4r^{3}}Q^{2}+ \frac{\Omega^{2}r}{4}\Lambda-\frac{\Omega^{2}}{4r}-\frac{\partial_{v}r\partial_{u}r}{r}
\]
\[
\partial_{u}\partial_{v}log(\Omega^{2}) = -2Re(D_{u}\phi\overline{\partial_{v}\phi})-\frac{\Omega^{2}Q^{2}}{r^{4}}+\frac{\Omega^{2}}{2r^{2}}+\frac{2\partial_{v}r\partial_{u}r}{r^{2}}
\]

Where equation (2.10) is modified slightly to include a nonzero contribution from $A_{u}$ as defined in equation (2.4) of [Ung22]. We also have an identical set of Raychaudhuri's equations as in the Minkowski case:

\[
\partial_{u}(\frac{\partial_{u}r}{\Omega^{2}}) = -r\frac{|D_{u}\phi|^{2}}{\Omega^{2}}
\]
\[
\partial_{v}(\frac{\partial_{v}r}{\Omega^{2}}) = -r\frac{|D_{v}\phi|^{2}}{\Omega^{2}}
\]

\section{\textbf{Extremality in dS/AdS spacetime}}

For de Sitter Reissner-Nordström black holes, we need to consider the full picture of extremality. This occurs when, for some horizon \( r_{+} \), we have \( f(r_{+}) = f'(r_{+}) = 0 \). For a given \( \Lambda > 0 \) and mass \( M \), the function \( f(r) \) is defined as:
\[
f(r) = 1 - \frac{\Lambda r^{2}}{3} - \frac{2M}{r} + \frac{Q^{2}}{r^{2}}. \label{eq:fr}
\]
From this, the critical point in the function occurs at a value of \( Q \) where the number of positive solutions (as a function of \( r_{+} \)) changes from 1 to 3. This marks the critical point of extremality.

\[
1 - \frac{\Lambda r^2}{3} - \frac{2M}{r} + \frac{Q^2}{r^2} = 0\\
- \frac{2\Lambda r}{3} + \frac{2M}{r^2} - \frac{2Q^2}{r^3} = 0.
\]

Solving these yields 

\[
+\frac{\Lambda r}{3} + \frac{2M}{r^2} - \frac{Q^2}{r^3} = \frac{1}{r}
\]

which has solutions at 

\[
r_{+} = \frac{\sqrt{2}}{2} \sqrt{\pm \frac{\sqrt{-4\Lambda Q^2 + 1}}{\Lambda} + \frac{1}{\Lambda}}
\]

Thus we have a set of extremal horizons, where the outermost extremal horizon is well-defined for $Q < \frac{1}{2\sqrt{\Lambda}}$. Any such solution satisfies $r_{+} < \sqrt{\frac{3}{\Lambda}}$, as required by the proof.\footnote{See: Appendix A}.

\section{\textbf{Sphere data sets in curved spacetime}}

For dS, AdS sphere data sets:
\[
ds^{2} = -f(r)dt^{2}+\frac{1}{f(r)^{2}}dr^{2}+r^{2}g_{S^{2}}
\]
\[
f(r) = 1\mp\frac{\Lambda r^{2}}{3}-\frac{2M}{r}+\frac{Q^{2}}{r^{2}}
\]

for $\pm$ corresponding to De Sitter and Anti-de Sitter space, respectively.
\\

We work in spacetimes which have either a cosmological horizon (dS) or a boundary at infinity (AdS), hence we limit the scope of the sphere data sets to be contained in some region of spherically-symmetric spacetime with radius $R_{BH} <R_{D} < R_{ch}$, where the final inequality is applicable in the de Sitter case. In either spacetime, since $\Lambda$ affects the lapse, we note that $\mathcal{G}$ contains $(f,g)$ such that the lapse can be normalized for a some region $R_{D}$ that does not extend to infinity. As in the original work, will also assume the gauge $\Omega = 1$; that is $\omega = 1$, $\omega_{u}^{i} = \omega_{v}^{i} = 0$ for $1 \leq i \leq k$. Every sphere data set then is gauge equivalent to a lapse normalized sphere data set. 

We redefine $\mathcal{G}$ as

\begin{center}
$\mathcal{G} = \{(f, g): f, g \in $ Diff$_{+}(\mathbb{R} \cap [0, R_{ch}]), f(0) = g(0) = 0\} \times C^{\infty}(\mathbb{R}\cap [0, R_{ch}])$\footnote{For a note on the limits of this $R_{D}$ in dS space, see Appendix A}
\end{center}

Furthermore, we define the sphere data sets for dS, AdS space as follows:
\\

\emph{$\textbf{Definition 4.1}$: de Sitter data. Let $k \in \mathbb{N}$, and $R > 0$. Then the unique lapse normalize sphere data set satisfying
\begin{enumerate}[itemsep=0pt]
\item $\rho$ = R
\item $\rho^{1}_{u} = -\frac{1}{2}(1-\frac{1}{3}\Lambda R^{2})$
\item $\rho^{1}_{v} = \frac{1}{2}(1-\frac{1}{3}\Lambda R^{2})$
\end{enumerate}}
is herein referred to as the dS (AdS) sphere data of radius R and vacuum energy $\Lambda > 0$ ($\Lambda < 0$). It is notated $D^{R, k}_{dS}$ ($D^{R, k}_{AdS}$).\footnote{For any sphere data set, all values listed are set to zero.}
\\

\emph{$\textbf{Definition 4.2}$: de Sitter-Schwarzschild data. Let $k \in \mathbb{N}$, and $0 \leq r_{sch} \leq R \leq r_{ch}$, where $r_{sch}$ is the dS-Schwarzschild radius and $r_{ch}$ is the Cosmological horizon. We assume to work in a regime for $M >> |\Lambda|$, such that $0 < r_{sch} < r_{ch}$. Then the unique lapse normalize sphere data set satisfying
\begin{enumerate}[itemsep=0pt]
\item $\rho$ = R
\item $\rho^{1}_{u} = -\frac{1}{2}(1-\frac{1}{3}\Lambda R^{2})$
\item $\rho^{1}_{v} = \frac{1}{2}(1-\frac{2M}{R}-\frac{1}{3}\Lambda R^{2})$\\
\end{enumerate}}
is the Schwarzschild-de Sitter sphere data set of mass M and cosmological constant $\Lambda > 0$, notated $D_{SdS}^{R, k, M}$
\\

\emph{$\textbf{Definition 4.3}$: AdS-Schwarzschild data. Let $k \in \mathbb{N}$, and $0 \leq r_{sch}' \leq R$, where $r_{sch}'$ is the apparent horizon of a $SAdS$ black hole with parameters $\Lambda, M$. Then the unique lapse normalize sphere data set satisfying
\begin{enumerate}[itemsep=0pt]
\item $\rho$ = R
\item $\rho^{1}_{u} = -\frac{1}{2}(1-\frac{1}{3}\Lambda R^{2})$
\item $\rho^{1}_{v} = \frac{1}{2}(1-\frac{2M}{R}-\frac{1}{3}\Lambda R^{2})$\\
\end{enumerate}}
is the Schwarzschild Anti-de Sitter sphere data set of mass M and cosmological constant $\Lambda < 0$, notated $D_{SAdS}^{R, k, M}$
\\

\emph{$\textbf{Definition 4.4}$: RNAdS data. Let $k \in \mathbb{N}$, and $1-\frac{2M}{r}+\frac{e^{2}}{r^{2}}-\frac{\Lambda r^{2}}{3}
$ with two roots, $r_{+}, r_{-} > 0$ corresponding to the inner and outer horizons of the black hole. Then the unique lapse normalize sphere data set satisfying
\begin{enumerate}[itemsep=0pt]
\item $\rho$ = $r_{+}$
\item $\rho^{1}_{u} = -1/2$
\item $\rho^{1}_{v} = 0$
\item $q = e$
\end{enumerate}}
is the Reissner-Nördstrom Anti-de Sitter sphere data set of mass M and cosmological constant $\Lambda$, notated $D_{RNAdS}^{R, k, M}$
\\

\emph{$\textbf{Lemma 4.1}$: Gauge Transformation dS/AdS identification. If $D \in D_{k}$ satisfies (for all $1 \leq i \leq k$):
\begin{enumerate}[itemsep=0pt]
\item  $\rho = R > 0$
\item $\rho_{u}^{1} < 0$
\item $\frac{1}{2}\rho(1+4\rho_{u}^{1}\rho_{v}^{1}) = M$
\item $Q = 0$
\item $\zeta_{u}^{i} = \zeta_{v}^{i} = 0$
\end{enumerate}}

Then it follows that $R$ is equivalent to $D_{SdS}^{R, k, M}$/$D_{SAdS}^{R, k, M}$ under $\mathcal{G}$.
\\

Similarly, we define gauge equivalence for $D_{RNAdS}^{R, k, M}$:
\\

\emph{$\textbf{Lemma 4.2}$: Gauge Transformation for $RNAdS$: If $D \in D_{k}$ satisfies (for all $1 \leq i \leq k$):
\begin{enumerate}[itemsep=0pt]
\item  $\rho = r_{+}$
\item $\rho_{u}^{1} < 0$
\item $\rho_{v}^{1} = 0$
\item $e = qM$ 
\item $\zeta_{u}^{i} = \zeta_{v}^{i} = 0$
\end{enumerate}}

The charge vanishes to all orders, so under a normalization of gauge potential and lapse we rescale $u \to \lambda u$ and $v \to \lambda^{-1}v$ to make $\rho_{u}^{1} = -\frac{1}{2}$. 
\\

\section{\textbf{k-regular gluing theorems in curved spacetime}}

We now trace the modifications to the proofs from Section 3.4 of [Ung22]. We first show the cases of gluing $D_{dS}^{R, k, M}$, $D_{AdS}^{R, k, M}$ to dS/AdS-Schwarzschild Cauchy data sphere sets $D_{SdS}^{R, k, M}$, $D_{SAdS}^{R, k, M}$ respectively. Theorems 5.1, 5.2 correspond to Theorem 2A in [Ung22], and Theorem 5.3 corresponds to Theorem 2B.
\\

\emph{$\textbf{Theorem 5.1}$: For any $k \in \mathbb{N}$, $0 < R < 2M$, and $\Lambda<< r_{D},$\footnote{See: dS gluing limit; Appendix A} the de Sitter sphere of radius $R$, $D_{dS}^{R, k, M}$, can be characteristically glued to a Schwarzschild-de Sitter event horizon sphere with mass $M$, $D_{SdS}^{R, k, M}$ with $C^{k}$ regularity, such that the solution globally satisfies the EMSCF system with SO(3) symmetry.}
\\

\emph{$\textbf{Theorem 5.2}$: For any $k \in \mathbb{N}$ and $0 < R < 2M$, the Anti-de Sitter sphere of radius $R$, $D_{AdS}^{R, k, M}$, can be characteristically glued to an Anti-de Sitter Schwarzschild event horizon sphere with mass $M$, $D_{ASdS}^{R, k, M}$ with $C^{k}$ regularity, such that the solution globally satisfies the EMSCF system with SO(3) symmetry.}
\\

\emph{$\textbf{Theorem 5.3:}$ For any $k\in \Bbb N$, $\mathfrak q\in [-1,1]$, and $ e\in \Bbb R\setminus\{0\}$, there exists a number $M_0(k, \mathfrak q, e)\ge 0$ such that if $M_f> M_0$, $0\le M_i\le \tfrac 18 M_f$, and $2M_i<R_i\le \tfrac 12 M_f$, then the Anti-de Sitter Schwarzschild sphere of mass $M_i$ and radius $R_i$, $D^\mathrm{S}_{M_i,R_i,k}$, can be characteristically glued to the Reissner-Nördstrom event horizon with mass $M_f$ and charge to mass ratio $q$, $D_{RNAdS}^{R, k, M}$, to order $C^k$. Furthermore, associated characteristic data can be chosen to have no spherically symmetric antitrapped surfaces.}\footnote{reduces to Theorem 2B [Ung22] in the case of $\Lambda = 0$}
\\

\section{$\textbf{modifications for Proof of Theorems 5.1, 5.2}$}

Condition 2 is different. It can be shown by considering the quantity $\partial_{v}(r\partial_{u}r)$, which can be shown via expanding and substitution of (16) to be:

\[
    \partial_{v}(r\partial_{u}r) = -\frac{1}{4}(1-r^{2}\Lambda)
\]

which propagates the sign of $r\partial_{u}(r)$ for small values of $\Lambda$, and all values of $\Lambda < 0$. See Appendix A for a specific discussion of this limit. By assumption, this holds and therefore $\partial_{u}r_{\alpha}(v) < 0$ on $[0,1]$.
\\

Immediately from the definition, the scalar field $\phi$ is odd in $\alpha$. Since $r_{\alpha}$ is determined by equation (4.18) which has only even powers of $\phi$, then $r_{\alpha}$ must be even in $\alpha$. We then combine wave equations to determine the wave equation for $r\phi$. Simple rearranging yields:

\[
\partial_{u}\partial_{v}(r\phi) = \phi(\frac{r\Lambda}{4}-\frac{1}{4r}-\frac{\partial_{u}(r)\partial_{v}(r)}{r})
\]

Since $\phi$ is odd in $\alpha$ and the expression depending on $r$ is even, the right-hand side of this expression is odd in $\alpha$. Thus it follows by inspection that $\partial_{u}(r\phi)$ is odd and therefore $\partial_{u}\phi$ is also odd. From here it follows from Proposition 6.2 that these parity rules hold for higher order derivatives as well. This can be seen in the transport equations for ingoing derivatives of r and $\Omega^{2}$ (which only contain even powers of $\phi$), and the analogous equations for $\partial_{u}^{i}\phi$ for $1 \leq i \leq k$ involve only odd powers of $\phi$. 
\\

\section{\textbf{modifications for proof of Theorem 5.3}}

The transverse derivative obtained in [Ung22] is different, but solved in the same manner: $\partial_u r(v;\alpha)$ is now determined by solving wave equation (2.9),
\[
w    \partial_v \partial_u r(v;\alpha)=- \frac{1}{4r(v;\alpha)^2}-\frac{\partial_ur(v;\alpha)\partial_vr(v;\alpha)}{r(v;\alpha)^2}+\Lambda\frac{\Omega^{2}r}{4}+\frac{Q(v;\alpha)^2}{4r(v;\alpha)^3},
\]
with initialization w.

Furthermore,
\[1-\frac{2M_i}{r(0;\alpha)}\geq 1-\frac{4M_i}{M_f}>0,\]
so
\[
\partial_ur(0;\alpha)<0.\label{ni-negative}
\]

Having initialized $\partial_u r $ at $v=0$, we determine $\partial_u r(v;\alpha)$ using the above wave equation, and we will now show that for  $ eM_f/ q$ sufficiently large, $\partial_u r(v;\alpha) <0$ for all $v\in [0,1]$.
\\

First, as in the proof of Theorem 5.1, we bound $\partial_{v}(r\partial_{u} r)$:
\[
    |\partial_v(r\partial_u r)|= |\frac 14 (1+\Lambda r^{2}-\frac{Q^2}{r^2})|\lesssim 1, 
\]
as
\[
    Q(v;\alpha) \leq  Q(1;\alpha)  =  q M_f  \lesssim r(v;\alpha),
   \]

The reminder of the proof is identical from here, with only a small modification to [Ung22] Lemma 4.9. Since we have redefined the gauge group $\mathcal{G}$ on a closed set $\mathbb{R} \cap [0, R_{ch]}$, the diffeomorphism

\[(\Re F^1,\Im F^1,..., \Re F^k,\Im F^k)\circ p_Q :S^{2k}\to \Bbb R^{2k},\]

must now map into the space $[0, R_{ch}]^{2k}$. However, since all components of im(F) are bounded\footnote{$im_{F}(S^{2k})$ is compact and therefore bounded in $\mathbb{R}^{2k}$}, they can be rescaled by a constant such that $L^\infty({\alpha}) < R_{ch}.$ Thus we still have such an $\alpha \in Q^{2k}$ where $F(\alpha)=0$. $D_{\alpha}(1) \simeq D_{RNAdS}^{R, k, M}$ which concludes the gluing construction, hence Lemma 4.9 still holds.

\section{\textbf{Teleology of a counterexample}}  

The close similarity to Kehle and Unger's work converges in this section, as both their theorems and the ones listed here serve to disprove the third law of black hole thermodynamics in their respective ambient spacetimes in the same manner. For the sake of clarity, we reframe the main corollary, and then follow a similar argument to Section 5.3 [Ung22].
\\

\emph{$\textbf{Corollary 8.1}$ For any $k \in \mathbb{N}$, $q \in [-1, 1] \backslash \{0\}$, and $e \in \mathbb{R} \backslash \{0\}$. Let $M_{0}(k, q, e)$ be defined as in Theorem 2B. Then for any $M \geq M_{0}$ there exist asymptotically $AdS$, spherically-symmetric Cauchy data for the EMSCF system, with $\Sigma \simeq \mathbb{R}^{3}$, and a regular center, such that the maximal future globally hyperbolic development $(\mathcal{M}^{4}, g, F, A, \phi)$ has the following properties:
\begin{enumerate}[itemsep=3pt, parsep = 0pt]
\item  All dynamical quantities are at least $C^{k}$-regular.
\item Null infinity $\mathcal{I}^{+}$ is complete.
\item The black hole region $(\mathcal{M} \backslash J^{-}(\mathcal{I}^{+}))$ is non-empty.
\item The Cauchy surface $\Sigma$ lies in the domain of outer communication $J^{-}(\mathcal{I}^{+}).$
\item The initial surface does not contain trapped surfaces
\item The spacetime does not contain antitrapped surfaces.
\item For sufficiently late advanced times $v \geq v_{0}$, the domain of outer communication is isometric to that of a RNAdS black hole with mass $M$ and mass-to-charge ratio $q$.
\end{enumerate}}

Using the general gluing methodology outlined above (as well as the results from Theorem 5.3), we can take a portion of empty AdS space identified with

\[
t+r \leq \frac{1}{2}M
\]
\[
t-r \geq -\frac{1}{2}M
\]

and glue to a RNAdS solution with the desired parameters; this allows for a complete future neighborhood of the event horizon. We now identify a spacelike curve $\Sigma$ connecting spacelike infinity $i^{0}$
in the exactly RNAdS region to the center, to the past of the cone $u = -1$. The curve $\Sigma$ can be chosen so the induced data on
it is asymptotically flat near $i^{0}$.

Completeness of null infinity $\mathcal{I}^{+}$ is inherited from the exact RNAdS solution, which can be seen from the fact that $\Sigma$ can be contained in $J^{-}(\mathcal{I}^{+}$. For the statement regarding trapped surfaces, see [Ung22] Proposition B.2, which follows identically in flat ambient space. Antitrapped spheres are also prevented by Raychaudhuri's equation, which propagates the sign of $\partial_{u}(r)$. 
\\

Since the global hyperbolic development of $(\mathcal{M}^{4}, g, F, A, \phi)$ lies in the causal future of the induced data, uniqueness of Maxwell Field global hyperbolic development guarantees the solution satisfies Corollary 1.
\\

\section{\textbf{the third law of black hole thermodynamics}}

In this section we analyze the behavior of extremal $AdS_{4}$ Reissner-Nördstrom black holes, and show that, under the definitions of extremality posed by [Isr92] and [Ung22], counterexamples exist to the third law in $AdS$ spacetime. 
\\

\emph{$\textbf{Theorem 9.1}$: For any $k \in \mathbb{N}$ and $e \in \mathbb{R} \backslash \{0\}$, there exist asymptotically flat, spherically symmetric Cauchy data $(\Sigma, g_{0}, k_{0}, e_{0}, B_{0}, \phi_{0}, \phi_{1})$ with $\Sigma \simeq \mathbb{R}^{3}$ and a regular center such that the maximal future hyperbolic development $(\mathcal{M}^{4}, g, F, A, \phi)$ has the following properties:
\begin{enumerate}[itemsep=0pt]
\item  All dynamical quantities are at least $C^{k}$-regular.
\item The spacetime and Cauchy data satisfies Corollary 1 for $q^{*} = 1$ and final mass $M_{f} \geq M_{0}(1, e, k)+8$.
\item The spacetime contains a double null rectangle of the form $\mathcal{R} = \{-2 \leq u \leq -1\} \cup \{1 \leq v \leq 2\}$ which is isomorphic to a Schwarzschild-AdS solution with $M = 1$.
\item The cone $\{u = -1\} \cup \mathcal{R}$ lies in the outermost horizon $\mathcal{A}'$ of the spacetime and is isometric to an appropriate portion of the r = 2 hypersurface in the Schwarzschild-AdS spacetime of mass 1.
\item The outermost apparent horizon $\mathcal{A}'$ is disconnected
\item $\mathcal{M} \backslash J^{-}(\mathcal{I}^{+})$ contains trapped surfaces for arbitrary advanced time.
\item For sufficiently late affine times $\tau_{1}, \tau_{0}; \tau_{1} > \tau_{0}$, there exists some neighborhood $\mathcal{N}$ of the apparent horizon $\mathcal{H}^{+}_{\tau \geq \tau_{1}}$ which guarantees $\mathcal{N} \backslash \mathcal{H}^{+}_{\tau \geq \tau_{0}}$ contains only strictly untrapped surfaces ($\partial_{v}r > 0$).
\end{enumerate}}

We begin by gluing an AdS cone to a Schwarzschild-AdS event horizon ($M = 1$) along $\{u = -1\}$. Then attach a double-null rectangle equipped with Schwarzschild-AdS data along $r = 2$ up to $v = 2$. Place $u = -2$ such that
\[
sup_{\{u = -2\} \cap \mathcal{R}} r = 2 + \epsilon \leq 3
\]

For $\epsilon$ sufficiently small, the first strip down to the center can be constructed as in the proof of Corollary 1. Let $M_{f} \geq M_{0}+8$ and extend the cone $u = -2$ to the future with vanishing scalar field until $r = \frac{1}{2}(M_{0}+8) > 3$. By Theorem 5.3, extremal Reissner-Nördstrom AdS data can be attached. The asymptotically flat spacelike curve (which attaches at $i^{0}$) is constructed as before.
\\

Then the maximal future global hyperbolic development $(\mathcal{M}^{4}, g, F, A, \phi)$ constrains the causal codomain of $\Sigma$ and satisfies the conclusions of the previous corollary. 
\\

$\mathcal{M}$ also contains trapped surfaces in any future neighborhood of $\{u = -1\} \cup \mathcal{R}$, as $\partial_{v}r = 0$ along $\{u = -1\} \cup \mathcal{R}$ and, from (32), it still holds that 
\[
    \partial_{u}(r\partial_{v}r) = -\frac{\Omega^{2}}{4}(1-r^{2}\Lambda)
\]

Which, for $\Lambda < 0$ guarantees this by definition. To show that trapped surfaces persist for arbitrary advanced time, Kehle and Unger's argument is identical:
\\

\emph{"...we invoke the general boundary condition of [Kom13]. If the $r = 0$ singularity $\mathcal{S}$ is empty, then the outgoing cone starting from one of
these trapped spheres terminates on the Cauchy horizon $\mathcal{CH}^{+}$ and the claim is clearly true by Raychaudhuri’s
equation. If $\mathcal{S}$ is nonempty, then every outgoing null cone which terminates on $\mathcal{S}$ is eventually trapped,
since $r$ extends continuously by zero on $\mathcal{S}$. Furthermore, $\mathcal{S}$ terminates at the Cauchy horizon $\mathcal{CH}^{+}$ or future timelike infinity $i^{+}$."}
\\

Unlike in the case of a flat background spacetime, we do not have the condition that there exists a neighborhood $U$ of $\mathcal{H}^{+}$ in $\mathcal{M}$ such that there are no trapped surfaces $\mathcal{S} \subset \mathcal{U}$. For any $p \in \mathcal{H}^{+}$ representing a sphere after the final gluing sphere. Then $r(p) = Q^{*}(p) = M_{f}$, $\partial_{v}r = 0$, and $\phi(p) = 0$. For $\Omega^{2} = 1$ parameterization along the ingoing null cone passing through p, equation (32) reads:
\[
\partial_{u}\partial_{v}r = -\frac{1}{4M_{f}} + \frac{M_{f}^{2}}{4M_{f}^{3}} +\frac{r\Lambda}{4} = \frac{M_{f}\Lambda}{4}
\]

Though, taking further derivatives reveals
\[
\partial_{u}^{2}\partial_{v}r = -\frac{2\partial_{u}rM_{f}}{M_{f}^{2}} + \partial_{u}(r)\frac{\Lambda}{4}\\
= \partial_{u}r(\frac{\Lambda}{4}-\frac{2}{M_{f}}) = 0
\]

Since $\phi$ has compact support and thus $\partial_{u}^{s}\phi(p) = 0$ for any s\footnote{up to $C^{k}$-regularity}. Thus, $\partial_{v}r$ expanded around $p$ yields

\[
\partial_{v}r_{u,v} = \frac{M_{f}\Lambda}{4}u-r|\partial_{u}\phi|^{2}(\frac{\Lambda}{8}-\frac{1}{M_{f}})u^{2}
\]

Reparameterizing into $u = \frac{1}{2}(t-r)$ guarantees similar quadratic behavior in t. Then, just as there is some $u_{0}$ such that $\partial_{v}r > 0$ for $u>u_{0}$. With the above affine transformation to $t$ we guarantee some $\tau_{0}$ with the same property.  

Then choosing some arbitrary $\tau_{1} > \tau_{0}$ guarantees that, for any $\tau > \tau_{1}$, There is some neighborhood $\mathcal{N}$ of the apparent horizon $\mathcal{H}^{+}_{\tau \geq \tau_{1}}$ which guarantees $\mathcal{N} \backslash \mathcal{H}^{+}_{\tau \geq \tau_{0}}$ contains only strictly untrapped surfaces, as in the sense of [Isr92] \& [Ung22]'s definition of extremality.
\\

Finally, the claim about the disconnectedness of the outermost apparent horizon $\mathcal{A}'$ now follows from the fact that $\mathcal{A}' \cap \mathcal{H}^{+}$ is one connected component of $\mathcal{A}'$ which does not contain the set $\{u = -1\} \cap \mathcal{R} \subset \mathcal{A}'$. 

\section{\textbf{Appendix A: an Upper limit on $\Lambda$ in de sitter spacetime}}

As mentioned in section 8, the $\Lambda > 0$ case requires we place limits on $r$. This is done to ensure that the sign of $\partial_{u}r$ is preserved along $v$. To find the limit on this, we begin with the wave equation: 

\[
\partial_{v}(r\partial_{u}r) = -\frac{1}{4}(1-r^{2}\Lambda)\nonumber 
\]

To solve for $\partial_{u}r(1)$, we use the identity:

\[
r\partial_{u}r(1) = r(0)\partial_{u}r(0) + \int_{0}^{1}\partial_{v}(r\partial_{u}r) dv\\
\]

By assumption, $r(0)\partial_{u}r(0) = -\frac{2}{R}(1-\frac{1}{3}\Lambda R^{2})$. Using (40), the above expression reduces to:

\[
r(1)\partial_{u}r(1) = \Lambda(\frac{2R}{3}+\frac{3R^{3}}{4})-\frac{8+R}{4R}
\]

Solving for $\partial_{u}r(1) > 0$ then algebraically yields the following inequality on Schwarzschild de-Sitter (SdS) gluing:

\[
\Lambda < \frac{3(8+R)}{R(8+9R^{2})}
\]

Where values of $\Lambda$ which satisfy the inequality allow for Schwarzschild-de Sitter manifold gluing.

We also have a Mass-dependent upper bound on $\Lambda$, which can be found by solving for $R_{D} = \sqrt{\frac{3}{\Lambda}} < f(r_{+})$, where $f$ is a function of $M, Q, \Lambda$. For a given mass M and constant $\Lambda$, we can find $r_{+}$ by varying Q.

\[
0 = 1- \frac{\Lambda r^{2}}{3}-\frac{2M}{r}+\frac{Q^{2}}{r^{2}}
\]
\[
r_{+} < \sqrt{\frac{3}{\Lambda}}
\]

\section{\textbf{Appendix B: EMSCF equations at conformal boundary in AdS spacetime}}

Because of the conformal boundary in AdS, we note that the EMSCF system in this regime demands special consideration. For stability of tachyonic modes of $\phi$ in AdS, we require that the solutions satisfy the Breitenlohner-Freedman bound at conformal infinity in order to be stable. 

At the conformal boundary, the EMSCF equations reduce to the asymptotic decay of $\phi$ and $A_{\mu}$, which regulate the interpretation in a holographic setup. The AdS system proposed in this note assumes that reflective boundary conditions hold, though the BF bound may imply that not all of the solutions generated by the method outlined above are stable. The model's reliance on smooth asymptotics requires compatibility between initial and boundary data, further imposing constraints on the scalar field mass and potential.
\\

\section{\textbf{Acknowledgments}}

First, I would like to recognize my advisor Mihalis Dafermos for his unbounded support and guidance through the entire process of writing this paper; his reassurance has been the keystone in my self-confidence, and his questions have been invaluable for my growth as a scientist. I would also like to extend my deep gratitude toward Ryan Unger for his willingness to motive and explain his work to an undergraduate—his insights are indispensable in nearly every section of this work. I am also grateful to Professor Pretorius for answering my tangential questions about black holes and Cauchy surfaces after class. Finally, I would also like to thank my friends, Paolo, Maguire, Luke, and Saarthak for their unrelenting encouragement.

\section{\textbf{References}}

1. [Isr92] W. Israel. “Thermodynamics and internal dynamics of black holes: Some recent developments”. Black hole physics. Springer, 1992, pp. 147–183.

2. [Kom13] J. Kommemi. “The global structure of spherically symmetric charged scalar field spacetimes”. Comm. Math. Phys. 323.1 (2013), pp. 35–106.

3. [OS39] J. R. Oppenheimer and H. Snyder. “On continued gravitational contraction”. Phys. Rev. (2) 56.5 (1939), pp. 455–459.

4. [Ros23] F. Rossetti. "Strong cosmic censorship for the spherically symmetric Einstein-Maxwell-charged-Klein-Gordon system with positive $\Lambda$: stability of the Cauchy horizon and $\mathcal{H}_{1}$ extensions" (2023). arXiv: 2309.14420.

5. [Ung22] R. Unger and C. Kehle. "Gravitational collapse to extremal black holes and the third law of black hole thermodynamics" (2022). arXiv: 2211.15742.

\end{document}